\documentclass[11pt,a4paper,english,superscriptaddress,aps]{revtex4}
\usepackage{amsmath,amssymb,graphicx}
\makeatletter
\usepackage{babel}

\usepackage{hyperref}

\newcommand{\be}{\begin{equation}}
\newcommand{\ee}{\end{equation}}
\newcommand{\bea}{\begin{eqnarray}}
\newcommand{\eea}{\end{eqnarray}}

\newcommand{\pa}{\partial}

\begin{document}

\title{Lorentz violation in the linearized gravity}

\author{A. F. Ferrari}
\author{M. Gomes}
\affiliation{Instituto de F\'\i sica, Universidade de S\~ao Paulo\\
Caixa Postal 66318, 05315-970, S\~ao Paulo, SP, Brazil}
\email{alysson,mgomes,ajsilva,jroberto@fma.if.usp.br}
\author{J. R. Nascimento} 
\affiliation{Instituto de F\'\i sica, Universidade de S\~ao Paulo\\
Caixa Postal 66318, 05315-970, S\~ao Paulo, SP, Brazil}
\affiliation{Departamento de F\'{\i}sica, Universidade Federal da Para\'{\i}ba\\
 Caixa Postal 5008, 58051-970, Jo\~ao Pessoa, Para\'{\i}ba, Brazil}
\email{jroberto,passos,petrov@fisica.ufpb.br}
\author{E. Passos}
\author{A. Yu. Petrov}
\affiliation{Departamento de F\'{\i}sica, Universidade Federal da Para\'{\i}ba\\
 Caixa Postal 5008, 58051-970, Jo\~ao Pessoa, Para\'{\i}ba, Brazil}
\email{jroberto,passos,petrov@fisica.ufpb.br}
\author{A. J. da Silva}
\affiliation{Instituto de F\'\i sica, Universidade de S\~ao Paulo\\
Caixa Postal 66318, 05315-970, S\~ao Paulo, SP, Brazil}
\email{mgomes,ajsilva@fma.if.usp.br}

\begin{abstract}
We study some physical consequences of the introduction of a Lorentz-violating modification term in the linearized gravity, which leads to modified dispersion relations for gravitational waves in the vacuum. We discuss two possible mechanisms for the induction of such a term in the Lagrangian. First, it is generated at the quantum level by a Lorentz-breaking coupling of the gravity field to a spinor field. Second, it appears as consequence of a particular modification of the Poisson algebra of the canonical variables, in the spirit of the so-called ``noncommutative fields approach''.
\end{abstract}

\maketitle

The idea that the quantization of gravity must somehow imply in the noncommutativity of spacetime at the Planck scale has a long history~\cite{Mead, Dopli}, and one of the most discussed issues concerns the fate of Lorentz symmetry in such a scenario. The first proposals for noncommutative spacetimes were carefully built to be compatible with Lorentz invariance~\cite{Snyder, Dopli}, an attitude still coherent with the very stringent bounds on Lorentz violation derived from the experiments~\cite{Mattingly:2005re}. Even so, a great amount of work have been done exploring the possibility of very small departures of Lorentz invariance as a source of observable signals of new physics. One of the most interesting possibilities is the modification of the dispersion relations governing the propagation of particles in the vacuum, which could generate outstanding effects in the spectrum of high energy cosmic rays~\cite{Coleman:1998ti}, among other consequences~\cite{Mag}.

The Lorentz violating standard model studied in~\cite{Colladay:1998fq} lists all types of interaction that, despite breaking the invariance of the theory under (particle) boosts/rotations, still preserve the gauge symmetry and the renormalizability of the standard model, thus furnishing a general background for the investigations about Lorentz violation. As for gravity theories, the incorporation of Lorentz violation is more delicate, specially when the violation is not spontaneous~\cite{Kostelecky,Kostel1}. It is interesting to look for alternative ways to introduce Lorentz violation in gravity, in such a way that simple physical effects can be drawn and discussed. This is the main objective of the present work. For the sake of simplicity, we consider the linearized gravity theory modified by one Lorentz violating term in the Lagrangian and consider the consequences of this modification. After that, we will discuss two distinct mechanisms for the generation of such a deformation, first as a quantum effect due to the coupling of gravity with a fermion in a Lorentz violating way, in the spirit of~\cite{Nasc1}, second as a consequence of a deformation of the canonical algebra~\cite{Gamboa1}, in the so-called ``noncommutative fields approach''. We will discuss the interesting features and the open issues in each mechanism. 

The starting point of our study is the Einstein-Hilbert action in the weak field approximation, also known as the Fierz-Pauli action (see f.e.~\cite{Veltman,DJT,GT}),
\be 
\label{lagr} 
S_{\rm FP}=\int d^4 x \Big(
\frac{1}{2}[\pa_{\lambda}h_{\mu\nu}\pa^{\mu}h^{\lambda\nu}-\pa_{\lambda}h\pa_{\mu}h^{\mu\lambda}]
+\frac{1}{4}[\pa_{\mu}h\pa^{\mu}h-\pa_{\lambda}h_{\mu\nu}\pa^{\lambda}h^{\mu\nu}]
\Big)\,.
\ee 

\noindent
Here the $h_{\mu\nu}$ is a second rank symmetric tensor characterizing (weak) metric fluctuations
($h_{\mu\nu}=g_{\mu\nu}-\eta_{\mu\nu}$, where $g_{\mu\nu}$ is the metric tensor of the curved space, $\eta_{\mu\nu}=diag(-+++)$ is the metric tensor of the flat space and $h=\eta^{\mu\nu}h_{\mu\nu}$ is the trace of $h_{\mu\nu}$). This theory is invariant under the gauge transformations
\bea 
\label{gauge} 
\delta
h_{\mu\nu}=\frac{1}{2}(\pa_{\mu}\xi_{\nu}+\pa_{\nu}\xi_{\mu})\,,
\eea 

\noindent
which are the linearized form of the diffeomorphism transformations, $\xi_{\mu}$ being the infinitesimal changes of the coordinates. The equations of motion look like
\bea
-\frac{1}{2}(\pa^{\lambda}\pa_{\mu}h_{\lambda\nu}+\pa^{\lambda}\pa_{\nu}h_{\lambda\mu})+\frac{1}{2}\eta_{\mu\nu}\pa_{\lambda}\pa_{\rho}h^{\lambda\rho}+\frac{1}{2}\pa_{\mu}\pa_{\nu}h+\frac{1}{2}\Box h_{\mu\nu}-\eta_{\mu\nu}\frac{1}{2}\Box h\,=\,0\,.
\eea

\noindent
The vanishing of the divergence of the left-hand-side of this equation is the linearized Bianchi identity.

We propose to modify Eq~(\ref{lagr}) by introducing an additional term $\Delta L$ in the Lagrangian,
\begin{equation}
\label{smod}
{\tilde S}_{\rm mod}\,=\,S_{\rm FP}+\int d^4 x \Delta L\,,
\end{equation}

\noindent
where
\bea
\label{newterm}
\Delta L\,=\,-2\epsilon^{\lambda\mu\nu\rho}\theta_{\rho}
h_{\nu\sigma}\pa_{\lambda}h_{\mu}^{\sigma}\,.
\eea

\noindent
Here, $\theta^{\rho}=(0,\theta^i)$, $i=1,2,3$, is a parameter for the Lorentz violation introduced in the theory, whose origin will depend on the mechanism inducing such a term in the action, as we will discuss later. For now, we have just to keep in mind that, despite written in a formally covariant way, $\Delta L$ is not a scalar, since $\theta^{\rho}$ is not a vector, but only a way to label a collection of three deformation parameters $\theta^i$. Also, the numerical coefficient in front of $\Delta L$ has been chosen for convenience. This particular form assumed by $\Delta L$ reproduces one of Lorentz-violating terms presented in~\cite{Kostelecky}, and the main motivations for its introduction are the physical consequences that will be drawn from it. 

The gauge transformations in Eq.~(\ref{gauge}) imply in the following variation of $\Delta L$,
\bea
\label{vardel}
\delta \Delta L\,=\,2\epsilon^{\lambda\mu\nu\rho}\theta_{\rho}\xi_{\nu}\pa_{\lambda}\pa_{\sigma}h^{\sigma}_{\mu}\,,
\eea 

\noindent
which does not vanish in general, so that the action ${\tilde S}_{\rm mod}$ is not gauge invariant. One possibility to remedy this problem is to implement a kind of Stueckelberg procedure (see~\cite{Ruegg:2003ps} for a review, or~\cite{Porrati:2002cp} for an application of this idea to the linearized massive gravity). We substitute $h_{\mu \nu} \rightarrow h_{\mu \nu} + \pa_\mu A_\nu + \pa_\nu A_\mu$ in the gauge non-invariant term $\Delta L$, thus obtaining,
\begin{equation}
\label{smod1}
S_{\rm mod}\,=\,S_{\rm FP}+\int d^4 x \left( \Delta L\,+\,\Delta L_A \right)\,,
\end{equation}

\noindent
where
\begin{equation}
\label{La}
\Delta L_A \, \equiv \, 2 \epsilon^{\mu \nu \lambda \rho}\,\theta_{\mu}
\left( \pa_\lambda A_\nu \Box A_\rho + 2 h_{\nu \sigma}\pa_\lambda \pa^\sigma A_\rho
\right) \,.
\end{equation}

\noindent
Then, one can check that the action $S_{\rm mod}$ is invariant under the gauge transformations Eq.~(\ref{gauge}) together with 
\begin{equation}
\label{gaugeA}
\delta A_\mu \,=\, -\xi_\mu \,.
\end{equation}

The most interesting aspect of the addition of the modification term in Eq.~(\ref{newterm}) is its consequences for the propagation of gravitational waves. To investigate this matter, we follow~\cite{Jackiw} and split the components of the metric fluctuation as follows,
\bea
\label{decomp}
h_{00}&=&n,\quad\, h_{0i}=\tilde{n}_i+\pa_i n_L,\nonumber\\
h_{ij}&=&\left(\delta_{ij}-\frac{\pa_i\pa_j}{\nabla^2}\right)\phi+\frac{\pa_i\pa_j}{\nabla^2}\chi+(\pa_i\tilde{\lambda}_j+\pa_j\tilde{\lambda}_i)+
\tilde{h}_{ij},
\eea
where the tilde denotes transversality, that is, $\pa_i \tilde{n}^i=\pa_i\tilde{\lambda}^i=\pa_i \tilde{h}^{ij}=0$. Besides that, the $\tilde{h}_{ij}$ is also traceless. Under a gauge transformation~(\ref{gauge}), these components transform as
\begin{eqnarray}
\label{compgauge}
\delta n = \dot{\xi}_0 \quad;\quad \delta n_L = \frac{1}{2}\left(\xi_0 + \frac{1}{\nabla^2}\pa_i \dot{\xi}_i \right)
\quad;\quad \delta \tilde{n}_i = \frac{1}{2}\left( \delta_{ij} - \frac{1}{\nabla^2} \pa_i \pa_j \right) \dot{\xi}_j \nonumber\\
\delta \chi = \pa_i \xi_i \quad;\quad \delta \phi = 0 \quad;\quad 
\delta \tilde{\lambda}_i = \frac{1}{2}\left( \delta_{ij} - \frac{1}{\nabla^2} \pa_i \pa_j \right) \xi_j  \quad;\quad   \delta \tilde{h}_{ij} = 0 \,.
\end{eqnarray}

Using the decomposition~(\ref{decomp}) in Eq.~(\ref{smod1}), we can derive the equations of motion for $\tilde{h}_{ij}$,
\begin{equation}
\label{eomh}
\frac{1}{2}\Box \tilde h_{ij} + 2 \left[ \theta_{ik} \pa_k \dot{ \tilde \lambda}_j +  \theta_{ik}\dot{\tilde h}_{kj} - \theta_{ik} \pa_k \tilde{n}_j + (i \leftrightarrow j) \right]
\, = \, 0 \,.
\end{equation}

\noindent
Here, we define an antisymmetric symbol $\theta_{ij}$ by means of
\begin{equation}
\label{ncvec}
\theta_{ij}=-\epsilon_{0ijk}\theta^k \,.
\end{equation}

\noindent
It is interesting to realize that $\Delta L_A$ do not contribute to the equations of motion of the transversal part of $h_{\mu \nu}$. Even if $\Delta L_A$ is essential to guarantee the gauge invariance of the total action, it follows directly from Eq.~(\ref{compgauge}) that Eq.~(\ref{eomh}) is gauge invariant, despite containing no contribution from the $A$ field. We can further simplify Eq.~(\ref{eomh}) by assuming a solution where $\tilde{h}_{ij}$ is the only non-vanishing field. One has to check whether this ansatz is consistent with the equations of motion of $\tilde \lambda$ and $\tilde n$, 
\begin{eqnarray}
G_{\tilde \lambda _j }[n,\tilde n, n_L, \phi, \chi, \tilde \lambda] + 4 \theta_{ik}\pa_k \dot{\tilde h}_{ij} = 0 \,, \nonumber\\
G_{\tilde n_j }[n,\tilde n, n_L, \phi, \chi, \tilde \lambda] - 4 \theta_{ik}\pa_k {\tilde h}_{ij} = 0 \,,
\end{eqnarray}

\noindent
which also involve $\tilde{h}_{ij}$. Here, $G$ means some homogeneous function of all components of $h_{\mu \nu}$ except  $\tilde{h}_{ij}$. We can safely set all other fields to zero if $\tilde{h}_{ij}$ satisfies
\begin{equation}
\theta_{ik}\pa_k {\tilde h}_{ij} \, = \, 0 \,.
\end{equation}

\noindent
We satisfy this condition by choosing $\theta^i = ( 0, 0, \theta / 4 )$, such that the only nonvanishing $\theta_{ij}$ are $\theta_{12}=-\theta_{21}=-\theta/4$, and considering a wave propagating in the $x_3$ direction. In this case, Eq.~(\ref{eomh}) can be solved by the ansatz $\tilde{h}_{ij} = H_{ij}e^{iq^\mu x_\mu}$, $q = (E, \vec p)$, and we end up with only two independent equations,
\begin{equation}
\Box Z+2i\theta\dot{Z}=0 \quad ; \quad \Box \bar{Z}-2i\theta\dot{\bar{Z}}=0 \,,
\end{equation}

\noindent
where $Z=H_{11}-iH_{12}$, $\bar{Z}=H_{11}+iH_{12}$. The corresponding dispersion relations are given respectively by
\bea
\label{disprel}
E&=&-\theta\pm\sqrt{{p}^2+\theta^2},\nonumber\\
E&=&\theta\pm\sqrt{{p}^2+\theta^2}\,,
\eea

\noindent
with $p=|\vec{p}|$. Thus we found that the dispersion relations are modified. One could say that the propagation of gravitational waves in the deformed theory displays a kind of birefringence phenomenon. For this configuration, there are two types of excitations with the velocities of propagation different from each other and from the speed of light, similarly to the propagation of electromagnetic waves in the noncommutative space~\cite{Jackiw1}. It must be stressed that these two different polarizations (and velocities) correspond to two ``circular'' polarizations with respect to ${\theta}^i$. The group velocity $\tilde{c}={dE}/{dp}={p}/{\sqrt{p^2+\theta^2}}$ is always less than the speed of light. On the other hand, the phase velocity $\hat{c}={E}/{p}$ can be superluminal for one of the polarizations. Some consequences of this modified dispersion relation for cosmology have been studied in~\cite{Cai:2007xr}.

At this point, we comment on some general properties of the deformed gravity theory~(\ref{smod1}). First of all, we note that the symmetry of positive- and negative-energy solutions in Eq.~(\ref{disprel}) indicates that C-symmetry is preserved. The whole action can be checked to be CPT invariant~\cite{Kostelecky,sheikh}. Concerning conservation laws, since there exists a preferable direction in the space-time, the angular momentum is no longer conserved. However, the energy-momentum tensor in the weak field approximation is still conserved since the background (Minkowsky) space-time is homogeneous (note that when the metric fluctuations are not small, the energy-momentum tensor is in general not conserved~\cite{Kostelecky}). Following the Noether theorem, we can write the energy-momentum tensor as
\bea
\label{Theta}
T_{\,a}^b&=&-\delta_{\,a}^b\Big[\frac{1}{2}(\pa_{\lambda}h_{\mu\nu}\pa^{\mu}h^{\lambda\nu}-\pa_{\lambda}h\pa_{\mu}h^{\mu\lambda})+
\frac{1}{4}(\pa_{\mu}h\pa^{\mu}h-\pa_{\lambda}h_{\mu\nu}\pa^{\lambda}h^{\mu\nu})\Big]+\nonumber\\&+&
\pa_{\nu}h^{\nu\lambda}\pa_ah^b_{\lambda}-\pa_a h\pa_{\nu}h^{\nu b}-\nonumber\\&-&2\delta_a^b\epsilon_{\mu\nu\lambda\rho}\theta^{\nu}
h^{\rho\sigma}\pa^{\lambda}h^{\mu}_{\sigma}+2\epsilon^{b\mu\rho\nu}\theta_{\mu}\pa_ah_{\nu\lambda}h_{\rho}^{\lambda}\,,
\eea

\noindent
where $a,b$ are indices in the local Lorentz frame. From Eq.~(\ref{Theta}) follows that $\partial_b T_{\,a}^b= 0$, that is, the energy-momentum tensor is modified by $\theta$-dependent terms although it remains conserved. Moreover, it is easy to check that $T^{0i}\neq T^{i0}$, which is a natural consequence of the Lorentz breaking in the theory~\cite{Jackiw}. As for the Bianchi identities, they are not satisfied in general, and this problem can be traced back to the general incompatibility between explicit Lorentz violation and Riemann-Cartan geometry, as discussed in~\cite{Kostelecky}. In this sense, a model with explicit Lorentz violation in gravity should be understood as a test model, where one can search for interesting phenomena arising in a simpler setting, while models with spontaneous breaking of the Lorentz symmetry will lead to more complete theories~\cite{Kostel1}.

Up to now, we have been discussing the consequences of the addition of the modification term in the linearized gravity action~(\ref{lagr}); from now on, we start looking for mechanisms to generate this term. The first idea is to couple the gravitational field to a fermion field by means of a Lorentz violating interaction (which should be one of the possible interactions described in~\cite{Colladay:1998fq}). A similar approach was used in~\cite{Nasc1} to generate a Chern-Simons term in the four dimensional linearized gravity. In our case, we consider the linearized gravity coupled to a Dirac field as follows
\bea
\label{secferm}
S[h, \bar\psi, \psi]\,=\,\int d^4x\left(\frac{1}{2} i\bar\psi\Gamma^\mu\stackrel{\leftrightarrow}{\partial}_\mu\psi+\bar\psi h_{\mu\nu}\Gamma^{\mu\nu}\psi-\bar\psi b_\mu\gamma^\mu\gamma_5 \psi+m\bar{\psi}\psi\right)+S_{\rm FP} \, ,
\eea

\noindent
with $\Gamma^\mu=\gamma^\mu-\frac12 h^{\mu\nu}\gamma_\nu$ and 
$\Gamma^{\mu\nu}=\frac12 b^\mu\gamma^\nu\gamma_5-\frac{i}{16}(\partial_\rho h_{\alpha\beta})\eta^{\beta\nu}\Gamma^{\rho\mu\alpha}$. The $b^\mu$ is a constant vector responsible for the Lorentz violation. 

\begin{figure}[ht]
\centering
\begin{tabular}{ccccc}
\includegraphics[width=0.15\textwidth]{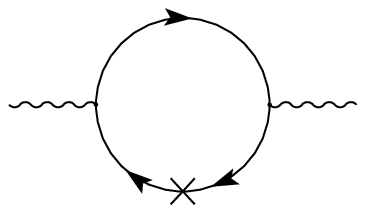} 
\includegraphics[width=0.15\textwidth]{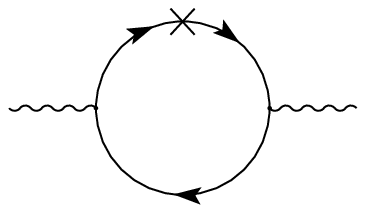} 
\end{tabular}
\caption{\label{fig1}One-loop contributions to the graviton two-point function. The cross in the fermion lines denotes a $\not \!b \gamma_5$ insertion.}
\label{oneloop}
\end{figure}

We first notice that the terms proportional to $\Gamma^{\mu\nu}$ in Eq.~(\ref{secferm}) will not contribute to $\Delta L$. The two-point vertex function of the graviton field receives contributions from the graphs in Fig.~\ref{fig1} (we use the Feynman rules that were explicitly indicated in~\cite{Nasc1}). The sum of these contributions, expanded up to the leading order in derivatives of the metric fluctuations, generates the following one-loop correction to the two-point vertex function of $h_{\mu\nu}$,
\bea
\Delta \Gamma^{(2)} \, = \, -i\epsilon^{\lambda\nu\rho\sigma}\int \frac{d^4k}{(2\pi)^4}\frac{k^{\mu}k^{\alpha}}{(k^2-m^2)^3}\Big[b_{\rho}(k^2+3m^2)-4k_{\rho}(b\cdot k)\Big]
h_{\mu\nu}\pa_{\sigma}h_{\alpha\lambda} \, .
\eea

\noindent
After the integration has been carried out with the use of the dimensional regularization, with $\varepsilon=4-d$, we obtain
\bea
\label{gamma2}
\Delta \Gamma^{(2)} \, = \,  \frac{m^2}{128\pi^2} \left(1+\frac{1}{\varepsilon}\right) \epsilon^{\lambda\nu\sigma\rho}b_{\rho} h_{\mu\nu}\pa_{\sigma}h^{\mu}_{\lambda} \, .
\eea

\noindent
Although $\Delta \Gamma^{(2)}$ correctly reproduces the structure of Eq.~(\ref{newterm}), its divergence forces us to include this term in the tree approximation with a coefficient adjusted to eliminate this divergence. From this viewpoint, this mechanism is not completely satisfactory to generate the $\Delta L$ term. In conclusion, the modification term can be generated in the gravity action by means of a Lorentz violating coupling with a fermion, but one has to deal with ultraviolet divergences. 

There is another possibility for the generation of a modified gravity theory similar to Eq.~(\ref{smod}), based on the deformation of the Poisson algebra of the canonical variables of the theory, with a consequent appearance of new terms in the classical action. The deformation used by us is inspired in the one presented in~\cite{Gamboa1}, where such a procedure was developed in the context of noncommutative fields theories~\cite{foot1} and later on was applied to electrodynamics and Yang-Mills theories~\cite{Gamboa2,Gamboa3}. As a result of the application of the noncommutative fields approach, the Hamiltonian of the theory and, as a consequence, the Lagrangian, is modified by new Lorentz-breaking terms. We will show that a term like the one in Eq.~(\ref{newterm}) will be produced in the (linearized) Einstein gravity as a consequence of the deformation of the canonical algebra. We remark that the possibility of generation of Lorentz-breaking terms for the gravity by perturbative corrections was shown in~\cite{Jackiw} (see also~\cite{Nasc1}). 

We start by reviewing the canonical quantization of the undeformed linearized gravity~(\ref{lagr}) (the non-linearized case was studied in~\cite{ADM} (see also~\cite{McKeon}); here we follow~\cite{GT} for the linearized case.). The first step is to construct the classical Hamiltonian. First, the indices of the Lagrangian are split into time and space ones. After some rearrangements, the Lagrangian corresponding to Eq.~(\ref{lagr}) takes
the form 
\bea 
\label{lagrold}
L_{\rm FP}=&-&\frac{1}{4}\dot{h}_{ii}\dot{h}_{jj}-\frac{1}{2}\pa_k
h_{ii}\pa_k h_{00}+\frac{1}{4}\pa_i h_{jj}\pa_i h_{kk}+\nonumber\\
  &+&\frac{1}{4}\dot{h}_{ij}\dot{h}_{ij}+\frac{1}{2}\pa_i h_{0j}\pa_i h_{0j}-\frac{1}{4}\pa_i h_{jk}\pa_i h_{jk}+\nonumber\\
  &+&\dot{h}_{ii}\pa_j h_{0j}-\frac{1}{2}\pa_i h_{kk}\pa_j h_{ij}+\frac{1}{2}\pa_i h_{00}\pa_j h_{ij}-\nonumber\\
  &-&\dot{h}_{ik}\pa_i h_{0k}-\frac{1}{2}\pa_i h_{0i}\pa_j h_{0j}+\frac{1}{2}\pa_i h_{jk}\pa_j h_{ik} \, .
\eea 

\noindent
Here the Latin indices stand for the pure space coordinates and $\dot f \equiv \pa_0 f$.
We see that the Lagrangian does not depend on the velocities $\dot h_{00}$ and $\dot h_{0i}$, so that $p^{0\mu}=\frac{\pa L}{\pa \dot{h}_{0\mu}}=0$. These are the primary constraints,
\bea 
\label{pric}
\Phi^{(1)}_{\mu} \, = \, p_{0\mu}\simeq 0 \, , 
\eea 

\noindent
which evidently commute with each other. The other momenta are given by 
\bea
\label{mom}
p_{ij} \, = \, \frac{\pa L}{\pa
\dot{h}_{ij}}=-\frac{1}{2}\dot{h}_{kk}\delta_{ij}+\frac{1}{2}\dot{h}_{ij}+\pa_k
h_{0k}\delta_{ij}-\frac{1}{2}(\pa_i h_{0j}+\pa_j h_{0i}) \, . 
\eea 

\noindent
Under the gauge transformations~(\ref{gauge}), they are transformed as
\bea
\delta p_{ij} \, = \, \frac{1}{2}(\delta_{ij}\pa_k\pa_k-\pa_i\pa_j)\xi_0 \, .
\eea

\noindent
The velocities are expressed from Eq.~(\ref{mom}) as 
\bea
\label{velo}
\dot{h}_{ij} \, = \, 2p_{ij}-p_{kk}\delta_{ij}+(\pa_i
h_{0j}+\pa_j h_{0i}) \, , 
\eea 

\noindent
and the canonical Hamiltonian density is given by
\bea 
\label{ham}
H&=&p^{\mu\nu}\dot{h}_{\mu\nu}(p)-L \nonumber\\ 
 &=&p_{ij}p_{ij}-\frac{1}{2}p_{kk}p_{ll}+\frac{1}{2}(\pa_i
h_{kk}\pa_j h_{ij}-\pa_i h_{jk}\pa_j h_{ik})+ \frac{1}{4}(\pa_i
h_{jk}\pa_i h_{jk}-\pa_i h_{jj}\pa_i h_{kk})-
\nonumber\\
&-& \frac{1}{2}h_{00}\big(\pa_i\pa_i
h_{kk}-\pa_i\pa_j h_{ij}\big) - 2h_{0j}\pa_i p_{ij} \,. 
\eea 

\noindent
Conservation of the primary constraints in Eq.~(\ref{pric}) requires $\{\Phi^{(1)}_{\mu},H\} \, \simeq \,  0$, where $\{\cdot,\cdot\}$ are the Poisson brackets, thus leading to the secondary constraints
\bea 
R_j\equiv\Phi^{(2)}_j \, = \, \pa_i
p_{ij}\simeq 0;\,\quad R_0\equiv \Phi^{(2)}_0 \, = \, \pa_l\pa_l
h_{kk}-\pa_i\pa_j h_{ij}\simeq 0. 
\eea 

\noindent
The $h_{00},h_{0i}$ are the Lagrange multipliers fields associated to these constraints. We note that the condition of conservation of the secondary constraints imply in only one new, tertiary constraint,
\bea
\label{f3}
\Phi^{(3)} \, = \, \{\Phi^{(2)}_0,H\}=\pa_i\pa_j p_{ij}\simeq 0 \, ,
\eea

\noindent
whereas $\{\Phi^{(2)}_i,H\}\equiv 0$.
The constraint $\Phi^{(3)}$ closes the Dirac algorithm since its Poisson bracket with the Hamiltonian is equal to zero (note that all the constraints in the theory are of first class). We note, however, that this constraint is really a spacial derivative of the $\Phi^{(2)}_j$, so its adding to the Hamiltonian with an arbitrary scalar multiplier $\lambda$ implies only in the replacement $h_{0j}\to h_{0j}+\pa_j\lambda$ in Eq.~(\ref{ham}). In fact, only the Lagrange multipliers associated to $\Phi^{(2)}_j$ are modified under such a replacement but not the constraints themselves. Therefore we will not consider the tertiary constraint henceforth (see~\cite{Ghalati:2007cf}).

The constraints $\Phi^{(2)}_0\equiv R_0$ and $\Phi^{(2)}_i\equiv R_i$
generate the gauge symmetry. For the quantization we must convert
the metric $h_{ij}$ and momenta $p_{ij}$ into operators whose commutation relations will be obtained from the classical Poisson brackets algebra,
\bea 
\label{poisson}
&&\left\{p_{ij}(\vec{x}),p_{kl}(\vec{y})\right\}=0,\nonumber\\
&&\left\{h_{ij}(\vec{x}),h_{kl}(\vec{y})\right\}=0,\nonumber\\
&&\left\{h_{ij}(\vec{x}),p_{kl}(\vec{y})\right\}=\frac{1}{2}\left(\delta_{ik}\delta_{jl}+
\delta_{jk}\delta_{il}\right)\delta(\vec{x}-\vec{y}) \, .
\eea 

\noindent
In this case the constraints $R_k,R_0$ generate, in the purely spacial sector (which is the physical sector), the gauge transformations,
\bea 
\label{gatra} 
\delta
h_{ij}&=&\{h_{ij},\Delta_{\xi}\}=\frac{1}{2}(\pa_i\xi_j+\pa_j\xi_i)\,,\nonumber\\
\delta
p_{ij}&=&\{p_{ij},\Delta_{\xi}\}=\frac{1}{2}(\delta_{ij}\pa_k\pa_k-\pa_i\pa_j)\xi_0 \,,
\eea 

\noindent
where 
\bea 
\label{delta} \Delta_{\xi}=-\int d^3 x
R_k(x)\xi_k(x)+\frac{1}{2}\int d^3 x R_0 \xi_0(x)
\eea 

\noindent
is the generator of gauge transformations.

Now, we follow the method described in~\cite{Gamboa2,Gamboa3}, and consider the simplest deformation of the classical Poisson bracket algebra as follows,
\bea 
\label{deform}
&&\left\{p_{ij}(\vec{x}),p_{kl}(\vec{y})\right\}=\theta_{ijkl}\delta(\vec{x}-\vec{y}),\nonumber\\
&&\left\{h_{ij}(\vec{x}),h_{kl}(\vec{y})\right\}=0,\nonumber\\
&&\left\{h_{ij}(\vec{x}),p_{kl}(\vec{y})\right\}=\frac{1}{2}\left(\delta_{ik}\delta_{jl}+
\delta_{jk}\delta_{il}\right)\delta(\vec{x}-\vec{y}),
\eea 

\noindent
where $\theta_{ijkl}$ is a $c$-number symbol possessing the
following symmetry:
$\theta_{1234}=\theta_{2134}=\theta_{1243}=-\theta_{3412}$.

For compatibility with the algebra in Eq.~(\ref{deform}), we modify the generators of the gauge transformations; instead of $\Phi^{(2)}_k=\pa_i p_{ik}$, we choose 
\bea 
\label{modgen}
R_k \, = \, \pa_i p_{ik}-\theta_{klnm}\pa_l h_{nm} \, .
\eea 

\noindent
The modified operator $\Delta_{\xi}$ implementing the gauge transformations in Eq.~(\ref{gatra}) takes the form
\be
\label{delta1}
\Delta_{\xi}=\frac{1}{2}\int d^3 x\Big\{(\pa_i \xi_k(x)+\pa_k
\xi_i(x)) \Big[ p_{ik}(x)+\theta_{rsik}h_{rs}(x)\Big]
-\xi_0(x)(\delta_{rs}\pa_l\pa_l-\pa_r\pa_s)h_{rs}(x)\Big\}\,.
\ee

\noindent
The modification of $\Delta_{\xi}$ (or, equivalently, of $R_k$) by an additive term proportional to $\theta_{rsik}$ implies in the modification of the Hamiltonian density: 
\bea
\label{newham}
H_{\rm new}&=&p_{ij}p_{ij}-\frac{1}{2}p_{kk}p_{ll}+ \frac{1}{2}(\pa_i
h_{kk}\pa_j h_{ij}-\pa_i h_{jk}\pa_j h_{ik})+ \frac{1}{4}(\pa_i
h_{jk}\pa_i h_{jk}-\pa_i h_{jj}\pa_i h_{kk})-\nonumber\\
&-&
\frac{1}{2}h_{00}\big(\pa_i\pa_i h_{kk}-\pa_i\pa_j h_{ij}\big) 
- 2h_{0j}(\pa_i p_{ij}-\theta_{jlnm}\pa_l h_{nm}) \,, 
\eea 

\noindent
which has been augmented by the term
\bea
\Delta H \, = \, 2h_{0j}\theta_{jlnm}\pa_l h_{nm} .
\eea 

\noindent
The velocities can be easily written as 
\bea
\dot{h}_{ij} \, = \, \{h_{ij},H_{\rm new}\}=2p_{ij}-p_{ll}\delta_{ij}+(\pa_i
h_{0j}+\pa_j h_{0i}) \, ,
\eea 

\noindent
reproducing Eq.~(\ref{velo}), from which the $p_{ij}$ are expressed in terms of velocities just like in Eq.~(\ref{mom}). The canonical conjugate momentum
of $h_{ij}$ (which should satisfy the commutation
relation $\{\hat{\pi}_{ij},\hat{\pi}_{kl}\}=0$) is $\hat{\pi}_{ij} \, = \, p_{ij}+\frac{1}{2}\theta_{klij}h_{kl}=p_{ij}$.

\noindent
The modified Lagrangian,
\bea
L_{\rm new} \, = \, \hat{\pi}_{ij}\dot{h}_{ij}-H_{\rm new} \, , 
\eea 
turns out to be
\bea
\label{lagrnew}
L_{\rm  new} &=& \frac{1}{2}(\pa_{\lambda}h_{\mu\nu} \pa^{\mu}h^{\lambda\nu}-\pa_{\lambda}h\pa_{\mu}h^{\mu\lambda})+
\frac{1}{4}(\pa_{\mu}h\pa^{\mu}h-\pa_{\lambda}h_{\mu\nu}\pa^{\lambda}h^{\mu\nu})-\nonumber\\&-&
2\theta_{jlnm}\pa_l h_{nm} +\frac{1}{2}\dot{h}_{ij}\theta_{klij}h_{kl},
\eea 

\noindent
differing from the initial Lagrangian by terms linear in $\theta$. 

One can study now particular cases of the general deformation~(\ref{deform}). The simplest case corresponds to 
\bea
\label{theta}
\theta_{ijkl}=\theta_{ik}\delta_{jl}+\theta_{il}\delta_{jk}+\theta_{jl}\delta_{ik}+\theta_{jk}\delta_{il} \, ,
\eea

\noindent
where $\theta_{ij}$ is a constant antisymmetric matrix to be related with the $\theta_i$ parameters in Eq.~(\ref{newterm}). In this case, the modified generators of gauge transformations take the form
\bea 
\label{modgen1}
R_k \, = \,\pa_i p_{ik}-2(\theta_{kn}\pa_l h_{ln}+\theta_{ln}\pa_l h_{kn}) \, ,
\eea 

\noindent
while the deformed Lagrangian turns out to be
\bea
\label{lagrnew1}
L_{\rm  new} &=& \frac{1}{2}(\pa_{\lambda}h_{\mu\nu} \pa^{\mu}h^{\lambda\nu}-\pa_{\lambda}h\pa_{\mu}h^{\mu\lambda})+
\frac{1}{4}(\pa_{\mu}h\pa^{\mu}h-\pa_{\lambda}h_{\mu\nu}\pa^{\lambda}h^{\mu\nu})-\nonumber\\&-&
4h_{0k}(\theta_{km}\pa_l h_{lm}+\theta_{lm}\pa_l h_{km})+\dot{h}_{ij}(\theta_{ki}h_{kj}+\theta_{kj}h_{ki}) \, ,
\eea 

\noindent
We can rewrite $\theta_{ij}$ in terms of three parameters $\theta^k$ by using Eq.~(\ref{ncvec}), thus obtaining $L_{\rm new}$ in the following form,
\bea
\label{lnew1}
L_{\rm new} &=& \frac{1}{2}(\pa_{\lambda}h_{\mu\nu}\pa^{\mu}h^{\lambda\nu}- \pa_{\lambda}h\pa_{\mu}h^{\mu\lambda})+
\frac{1}{4}(\pa_{\mu}h\pa^{\mu}h-\pa_{\lambda}h_{\mu\nu}\pa^{\lambda}h^{\mu\nu})+\nonumber\\&+&
2\epsilon^{\mu\nu\lambda\kappa}\theta_{\mu}
h_{\nu\sigma}\pa_{\lambda}h_{\kappa}^{\sigma}+
2\epsilon_{ijk}\theta_k h_{0i}\left(2\pa_l h_{lj}+2\pa_jh_{00}-\dot{h}_{0j}\right)\,,
\eea 

\noindent
where $\theta^{\mu} = (0, \theta^i)$. We notice the appearance of the term $\Delta L$, introduced in Eq.~(\ref{newterm}), which now is a consequence of a deformation of the canonical algebra of fields by a ``noncommutativity'' tensor $\theta_{ij}$. One should keep in mind that the parameters $\theta^i$ are not components of a vector, since Eq.~(\ref{ncvec}) is not covariant. 

Both $\Delta L$ and the last term in the action Eq.~(\ref{lnew1}), proportional to $h_{0i}$, are not gauge invariant. Contrarily to what happens in a spin-one gauge theory~\cite{Gamboa2,Gamboa3}, the noncommutative fields approach here conflicts with the gauge invariance. After the deformation of the algebra and the modification in the form of the constraints that generate the gauge transformations, we find that the modified constraints in Eq.~(\ref{modgen1}) are not first class any longer. 

One may wonder whether a more complicated deformation could improve matters here, and indeed this is the case. Consider the Poisson bracket between the deformed constraints in Eq.~(\ref{modgen}),
\bea
\label{prob}
\{R_k(\vec{x}),R_{k^\prime}(\vec{y})\} &=&\{\pa_i p_{ik}(\vec{x})-\theta_{klnm}\pa_l h_{nm}(\vec{x}),
\pa_{i^\prime} p_{i^\prime k^\prime}(\vec{x})-\theta_{k^\prime l^\prime n^\prime m^\prime }\pa_{l^\prime} h_{n^\prime m^\prime}(\vec{y})\}\nonumber\\& = &\theta_{iki^\prime k^\prime} \pa_i\pa_{i^\prime}\delta(\vec{x}-\vec{y})\,.
\eea
Notice that, in the spin-one models~\cite{Gamboa2,Gamboa3,NPR}, the deformation parameter $\theta$ carries only two indices and is antisymmetric, so that $\theta_{ik}\pa_i\pa_k\delta(\vec{x}-\vec{y})\equiv 0$. In our case, the algebra~(\ref{prob}) is again first-class if
\bea
\theta_{ijkl}=\theta_{ik}\tilde{\delta}_{jl}+\theta_{il}\tilde{\delta}_{jk}+\theta_{jl}\tilde{\delta}_{ik}+\theta_{jk}\tilde{\delta}_{il},
\eea

\noindent
where
\bea
\tilde{\delta}_{ij}=\delta_{ij}-\frac{\pa_i\pa_j}{\nabla^2}.
\eea

\noindent
Since $\tilde{\delta}_{ij}\pa_j=0$, it follows that $\theta_{iki^\prime k^\prime}\pa_i\pa_{i^\prime}\delta(\vec{x}-\vec{y})=0$ and the deformation of the canonical algebra yields a first-class system. The corresponding modification $\Delta L$ induced in the Fierz-Pauli Lagrangian~(\ref{lagr}) takes the form
\bea
\Delta L=2\theta_{ki}\dot{h}_{ij}\left(\delta_{lj}-\frac{\pa_l\pa_j}{\nabla^2}\right)h_{kl}
-4h_{0j}\theta_{ln}\left(\delta_{jm}-\frac{\pa_j\pa_m}{\nabla^2}\right)\pa_lh_{nm}.
\eea

\noindent
The appearance of the transversal projector in $\Delta L$ is reminiscent of the work~\cite{Dvali:2005ws}. It is clear that this $\Delta L$ possesses restricted gauge invariance under the transformations~(\ref{gauge}) with $\xi_i$ longitudinal, that is, those gauge parameters satisfying $(\delta_{jm}-\frac{\pa_j\pa_m}{\nabla^2})\xi_m=0$. From Eqs.~(\ref{compgauge}), we obtain the restricted gauge transformation for the components of $h_{\mu \nu}$ as,
\begin{align}
\label{newgauge}
\delta n = \dot{\xi}_0 \quad;\quad \delta n_L = \frac{1}{2}\left(\xi_0 + \frac{1}{\nabla^2}\pa_i \dot{\xi}_i \right)
\quad;\quad \delta \tilde{n}_i = 0  \nonumber\\
\delta \chi = \pa_i \xi_i \quad;\quad \delta \phi = 0 \quad;\quad 
\delta \tilde{\lambda}_i = 0 \quad;\quad   \delta \tilde{h}_{ij} = 0 \,.
\end{align}

The price is that the modified Lagrangian now contains non-local terms. However, it can be checked that the gauge symmetry~(\ref{newgauge}) allows us to go to the gauge $\pa_m h_{mn}=0$, where this non-local terms vanish, and $\Delta L$ reduces to
\bea
\label{newdelta}
\Delta L=2\theta_{ki}\dot{h}_{ij}h_{kj}-4h_{0m}\theta_{ln}\pa_lh_{nm}.
\eea

\noindent
One still has the freedom to set $h_{0m}=0$ by using the remaining gauge symmetry. This modification, yet not exactly equal to the one in Eq.~(\ref{smod1}), clearly leads to the same dispersion relations we studied before for the transversal-traceless part of $h_{\mu \nu}$. This is an example of a more subtle application of the noncommutative fields approach that solves, at least partially, the incompatibility with gauge symmetry for spin-two gauge theories we pointed out earlier. 

In this paper, we studied the introduction of a Lorentz violating term $\Delta L$ in the linearized gravity. Gauge symmetry is violated by such a term, but it can be restored using a Stueckelberg procedure, with the addition of a vector field $A$ to the modified Fierz-Pauli action. As an interesting physical consequence of such a deformation, we studied the propagation of gravitational waves, and found a kind of ``birefringence'' phenomenon in the vacuum. Next, we looked for mechanisms that could generate such a term in the linearized gravity action. First, the Lorentz violating coupling with a Dirac fermion does generate a $\Delta L$ term in the quantum effective action, but one has to face ultraviolet divergences. We have also shown that a deformation of the canonical algebra of fields, in the spirit of the ``noncommutative fields approach'' of~\cite{Gamboa1} does lead to the proposed term, together with additional terms in the action. Differently to the case of spin-one gauge theories~\cite{Gamboa2,Gamboa3}, here the deformation do not preserve the gauge invariance. This problem was at least partially solved by a more complicated deformation leading to a theory with a restricted gauge symmetry. The dispersion relations obtained from such a deformation are the same as the one induced by the addition of $\Delta L$ to the original Lagrangian. 

As a final remark, we note that in the three-dimensional Einstein gravity, a term structurally similar to $\Delta L$ would not violate Lorentz symmetry, since instead of $\epsilon^{\lambda\mu\nu\rho}\theta_{\rho}$, the factor $\epsilon^{\lambda\mu\nu}\theta$ would arise, with $\theta$ a scalar noncommutativity parameter. In three dimensions, $\Delta L$ is structurally similar to the usual Chern-Simons term~\cite{DJT} (which has arisen using the noncommutative fields approach in~\cite{NPR}).

\vspace{1cm}

{\bf Acknowledgements.} A. Yu. P. is grateful to R. Jackiw for useful discussions and to V. A. Kostelecky for some criticism on the manuscript. This work was partially supported by Funda\c{c}\~{a}o de Amparo \`{a} Pesquisa do Estado de S\~{a}o Paulo (FAPESP) and Conselho Nacional de Desenvolvimento Cient\'{\i}fico e Tecnol\'{o}gico (CNPq). The work by A. F. F. has been supported by FAPESP, project 04/13314-4. The work by A. Yu. P. has been supported by CNPq-FAPESQ DCR program, CNPq project No. 350400/2005-9.


\vspace{2cm}

\begin{thebibliography}{50}
\bibitem{foot1} This proposal became known as the ``noncommutative fields approach'', even if the space-time coordinates continue to be commutative, unlike other popular approaches of noncommutative field theories, such as the one based on the Moyal product~\cite{SW}, coherent states~\cite{Spa} and models with dynamical noncommutativity parameter~\cite{Morita} (see also~\cite{Riv,Bra}).
\end{thebibliography}
\end{document}